\documentclass[aps,prd,reprint,superscriptaddress,preprintnumbers,%
showpacs,nofootinbib]{revtex4-1}

\usepackage[colorlinks, pdfborder={0 0 0}, plainpages=false]{hyperref}
\usepackage[utf8]{inputenc}
\usepackage{graphicx}
\usepackage{amsmath}
\usepackage{amssymb}
\usepackage{acronym}

\newcommand{\AEIHannover}{Max Planck  Institute for Gravitational Physics
(Albert Einstein Institute), Callinstr.~38, 30167 Hannover, Germany}
\newcommand{\UniHannover}{Leibniz Universit\"at Hannover, D-30167 Hannover, Germany}

\newcommand{\cita}{Canadian Institute for Theoretical
    Astrophysics, 60 St.~George Street, University of Toronto,
    Toronto, ON M5S 3H8, Canada}

\newcommand{\cardiff}{School of Physics and Astronomy, Cardiff University, The
Parade, Cardiff, CF24 3AA, United Kingdom}

\newacro{BH}{black hole}
\newacro{BBH}{binary black hole}
\newacro{NR}{numerical relativity}
\newacro{GW}{gravitational wave}
\newacro{IMR}{Inspiral-Merger-Ringdown}
\newacro{NS}{neutron star}
\newacro{BNS}{Binary Neutron Star}
\newacro{LSC}{LIGO Scientific Collaboration}
\newacro{LVC}{LIGO-Virgo Collaboration}
\newacro{LIGO}{Laser Interferometric Gravitational-Wave Observatory}
\newacro{aLIGO}{Advanced LIGO}
\newacro{Adv}{Advanced Virgo}
\newacro{EM}{Electromagnetic}
\newacro{PN}{Post Newtonian}
\newacro{EOB}{Effective-One-Body}
\newacro{MSA}{multiple scale analysis}
\newacro{SUA}{shifted uniform asymptotics}
\newacro{SPA}{stationary phase approximation}
\newacro{SNR}{signal-to-noise ratio}
\newacro{PSD}{power spectral density}

\begin{document}

\preprint{LIGO-P1800251}

\title{ Phenomenological model for the gravitational-wave signal
from precessing binary black holes with two-spin effects }

\author{Sebastian Khan}
\affiliation\AEIHannover
\affiliation\UniHannover

\author{Katerina Chatziioannou}
\affiliation\cita

\author{Mark Hannam}
\affiliation\cardiff

\author{Frank Ohme}
\affiliation\AEIHannover
\affiliation\UniHannover

\date{\today}

\begin{abstract}

The properties of compact binaries, such as masses and spins, are imprinted in the
gravitational-waves they emit and can be measured using parameterised waveform models.
Accurately and efficiently describing the complicated \emph{precessional} dynamics of the various
angular momenta of the system in these waveform models is the object of active investigation.
One of the key models extensively used in the analysis of LIGO and Virgo data
is the \emph{single-precessing-spin} waveform model {\tt IMRPhenomPv2}.
In this article we present a new model {\tt IMRPhenomPv3} which includes
the effects of two independent spins in the precession dynamics.
Whereas {\tt IMRPhenomPv2} utilizes a single-spin frequency-dependent post-Newtonian rotation to describe precession effects,
the improved model, {\tt IMRPhenomPv3}, employs a double-spin rotation that is based on
recent developments in the description of precessional dynamics.
Besides double-spin precession, the improved model benefits from a more accurate
description of precessional effects.
We validate our new model against a large set of
precessing numerical-relativity simulations.
We find that {\tt IMRPhenomPv3} has better agreement
with the inspiral portion of precessing binary-black-hole simulations
and is more robust across a larger region of the parameter space than {\tt IMRPhenomPv2}.
As a first application we analyse, for the first time, the gravitational-wave event GW151226
with a waveform model that describes two-spin precession. Within statistical uncertainty our results are consistent
with published results.
{\tt IMRPhenomPv3} will allow studies of the measurability of individual spins of binary black holes using GWs
and can be used as a foundation upon which to build further improvements, such as
modeling precession through merger, extending to higher multipoles, and including tidal effects.
\end{abstract}


\maketitle

\section{Introduction}

Binary systems of compact bodies such as \acp{NS} and \acp{BH}
generate \acp{GW} that are detectable by second-generation ground based interferometers.
The energy carried away by \acp{GW} causes the orbit to decay
and the binary to merge on astrophysical time scales.
These tiny ripples in spacetime propagate almost completely unaffected
through the Universe and have the source properties imprinted
in the gravitational waveform.
In August 2017 the second observing run (O2) of the detectors aLIGO~\cite{TheLIGOScientific:2014jea} and
Virgo~\cite{TheVirgo:2014hva} ended, and to date O1 and O2 have resulted in the publication of 6 likely \ac{BBH}~\cite{2016PhRvL.116f1102A,2016PhRvL.116x1103A,2017PhRvL.118v1101A,Abbott:2017oio,Abbott:2017gyy} merger events, and the joint
GW-electromagnetic observation of a pair of merging \acp{NS}~\cite{TheLIGOScientific:2017qsa,GBM:2017lvd}.

Detection and characterisation of \ac{GW} signals is carried out by
a suite of software pipelines that analyse detector data using a variety of
analysis methods~\cite{PhysRevD.95.042001,Canton:2014ena,Usman:2015kfa,Nitz:2017svb,PhysRevD.93.042004}.
Matched-filter-based analyses, used
in the search for and parameter estimation~\cite{Cutler:1994ys,Veitch:2014wba,TheLIGOScientific:2016wfe} of \ac{GW} signals
from compact binaries require accurate and computationally-inexpensive models for the \ac{GW} signals,
which are used as \emph{templates}.
The need for computationally tractable analyses is best satisfied by frequency-domain models.
Moreover, the accuracy of \ac{GW} signal models and the quantity and quality of the physical effects
they include impact both the types of sources \ac{GW} analyses are sensitive to
and the fidelity of the conclusions we draw about their properties.

One such physical effect that has been at the center of waveform-modeling efforts in the recent years is spin-precession:
when the binary components' spin angular momenta are misaligned with the orbital angular momentum of the binary, spin-orbit and
 spin-spin interactions cause the binary orbit to change orientation in space, resulting in modulations in the observed signal amplitude
 and phase~\cite{PhysRevD.49.6274,PhysRevD.52.821}. Modeling these modulations is a challenging task, especially for systems with unequal-mass components
 that are observed from close to the binary plane, since this case requires accurate modeling of both spin-precession
 effects and higher modes beyond the dominant $(\ell,|m|)=(2,2)$ multipoles.

Numerous models have been developed in the recent years with the goal of
providing a complete coverage across the \ac{BBH} parameter space containing
all relavent physical effects required for the current sensitivity of
\ac{GW} detectors.

The effective-one-body--numerical-relativity (EOB-NR) model, {\tt SEOBNRv3}~\cite{PhysRevD.89.084006},
built upon the non-precessing model of~\cite{Taracchini:2013rva},
is a time-domain two-spin precessing model. It provides the
$(\ell,|m|) \in ((2,2),(2,1))$ multipoles in the co-precessing frame
although only the $(2,2)$ is calibrated to \ac{NR} data.
This model has been shown to accurately model the $\ell=2$ multipoles
from precessing \acp{BBH}~\cite{PhysRevD.95.024010}, however,
as it requires the integration of a set of coupled ordinary differential
equations it incurs a large computational cost.

The \emph{phenomenological} (phenom) waveform models are typically developed
and constructed in the frequency-domain and the precessing model {\tt IMRPhenomP}
was presented in~\cite{Hannam:2013oca}. {\tt IMRPhenomP} is built upon the
non-precessing model of~\cite{PhysRevD.82.064016} was later upgraded to
{\tt IMRPhenomPv2}
which used the non-precessing model of~\cite{PhysRevD.93.044006,Khan:2015jqa}.
Despite its computational efficiency {\tt IMRPhenomP}
is limited to single-spin effects,
though this choice was shown to not lead to appreciable biases
in the characterization of the first BBH signal~\cite{Abbott:2016izl,Abbott:2016wiq}.
A time-domain phenom model for precessing \acp{BBH}
was presented in~\cite{1742-6596-243-1-012007}, however, it is restricted to
equal-mass \acp{BBH} with spin magnitudes up to $0.6$.

Recently work has been done to develop models for multipoles beyond the
dominant $(\ell,|m|)=(2,2)$ multipole.
The first of such models were the non-spinning EOB-NR model of~\cite{PhysRevD.84.124052}
and a non-spinning phenom model was presented in~\cite{Mehta:2017jpq}.
The first extension of these higher multipole modes into spinning non-precessing
\acp{BBH} has been acomplished both by the phenom~\cite{PhysRevLett.120.161102}
and \ac{EOB}~\cite{Cotesta:2018fcv} approaches.

An alternative approach to waveform modeling has
 been successful in building \emph{surrogate} models~\cite{PhysRevX.4.031006} that
interpolate \ac{GW} data from \ac{NR} simulations. The surrogate model of~\cite{PhysRevD.96.024058}
is a fully precessing, time-domain model containing $\ell \leqslant 4$
multipoles, however, it is restricted to systems where the ratio of the components' masses is less than two.

In this paper we take another step towards a computationally-tractable \ac{GW} model that accurately models
generic BBH systems by introducing the frequency-domain fully-precessing phenomenological
model, {\tt IMRPhenomPv3}, that describes the inspiral, merger, and ringdown phases of
spinning \ac{BBH}s.
Our upgrade from {\tt IMRPhenomPv2} to {\tt IMRPhenomPv3} hinges on a novel closed-form solution to the differential equations that describe precession including the effects of radiation reaction
in the \ac{PN} regime where the binary components are well-separated \cite{Chatziioannou:2017tdw,Chatziioannou:2016ezg}.
This closed-form solution to the precession equations including the effects of radiation reaction has been shown to accurately
describe precession for systems with generic masses and spins \cite{Chatziioannou:2017tdw}.
Moreover, since it is an analytic frequency-domain model, it is also amenable to the reduced-order-quadrature method to greatly accelerate likelihood
evaluations for fast parameter estimation~\cite{Smith:2016qas}.
In addition, {\tt IMRPhenomPv2} exhibits non-physical precession behaviour for some high mass-ratio,
anti-aligned spin configurations, as discussed in Sec.~\ref{sec:v2issues}, but this behaviour is not observed in in
{\tt IMRPhenomPv3}.

We validate our new model by comparing against a
large set of precessing \ac{NR} waveforms.
We find that the improved treatment of the inspiral
improves the accuracy of the model for measurement of low-mass systems in LIGO-Virgo data.
Finally, as a first application and demonstration of our model's readiness we perform a Bayesian parameter
estimation analysis (using {\tt LALInference}~\cite{Veitch:2014wba}) on the
\ac{GW} event GW151226~\cite{2016PhRvL.116x1103A}. This is a $\sim$22$ \, M_\odot$ \ac{BBH} signal where at
least one of the \acp{BH} is measured to have a dimensionless spin
magnitude of $\gtrsim 0.2$ at the 99\% credible level. The presence of spin and the low total
mass, which implies a large number of \ac{GW} cycles ($\sim$55) measurable by the
detector, makes this an ideal candidate to look for evidence of
precession. This is the first analysis of GW151226 with a \ac{IMR}
precession model with the full 4 degrees of freedom coming from precession.

\section{Building the new model}

In this section we describe the construction of {\tt IMRPhenomPv3} and highlight the improvements compared to {\tt IMRPhenomPv2}.

\subsection{Precessing BBH phenomenology}

A quasicircular \ac{BBH} system can be parameterised
by only 7 intrinsic parameters;
the mass-ratio $q=m_1/m_2$\footnote{We use the convention $m_1\geqslant m_2$.}
and 6 spin angular momenta $\vec{S}_1$ and $\vec{S}_2$. In addition,
the total mass $M = m_1 + m_2$
can be factored out and systems with different total masses
can be obtained with appropriate scaling of $M$.
For each \ac{BBH} we can define a Newtonian orbital angular momentum
$\vec{L}$, which is perpendicular to the instantaneous
orbital plane and a total angular momentum $\vec{J} = \vec{L} + \vec{S}_1 +
\vec{S}_2$.

We classify \ac{BBH} systems into different categories according to their spin. \emph{Non-precessing} systems
have \ac{BH} spins that are either aligned- or anti-aligned
with $\vec{L}$.
Systems with spins (anti-)aligned with $\vec{L}$ plunge and merge
at (larger) smaller separations compared to nonspinning binaries, shifting the merger and ringdown part of the signal to (lower) higher \ac{GW} frequencies.
\emph{Precessing} \ac{BBH} are systems with arbitrary \ac{BH} spin orientation. Interactions between the \ac{BH} spins and $\vec{L}$ introduce
a torque on the orbital angular momentum causing it to precess around the (almost constant)
direction of the total angular momentum.
Special sub-categories of precessing systems include
\emph{simple-} and \emph{transitional-} precession binaries \cite{PhysRevD.49.6274, PhysRevD.52.821}.
Most binaries undergo simple precession, while transitional precession occurs when
$\vec{J} = \vec{L} + \vec{S}_1 + \vec{S}_2 \approx 0$~\cite{PhysRevD.49.6274}.

\subsection{Modeling precessing BBHs}

The complicated phenomenology of generically precessing BBHs makes waveform modeling especially challenging.
A significant advance was achieved when
it was observed that the \ac{GW} signal from precessing binaries is simplified when
observed in a frame that is adapted to the precessional motion of the binary \cite{PhysRevD.84.024046, PhysRevD.84.124002}.
In this non-inertial (co-precessing) frame the $z$-axis approximately tracks the orientation of the orbital plane.
When the waveform is transformed into this frame
it closely mimics the signal of the equivalent \ac{BBH} system that has the
spin components
perpendicular to $\vec{L}$ set to zero~\cite{PhysRevD.84.024046,PhysRevD.88.024040}.
This is a consequence of the approximate decoupling between the components of
spin parallel and perpendicular to $\vec{L}$; the former influences the rate of
inspiral and the latter drives the precessional motion~\cite{PhysRevD.49.6274,PhysRevD.84.024046,Buonanno:2002fy}.

This observation led to a method for building models for the gravitational
waveform from generic precessing \acp{BBH} by first constructing a model for
the gravitational waveform produced by the equivalent non-precessing \acp{BBH}
and then introducing precession through a time (or frequency) dependent
rotation of the signal derived from the orbital dynamics~\cite{PhysRevD.86.104063}.
Colloquially this procedure is denoted ``twisting-up'' a non-precessing
model to produce a precessing model~\cite{PhysRevD.86.104063}.

Beginning with an inertial frame that is aligned with the total angular momentum
at some reference frequency $\hat{z} = \hat{J}$ we describe the orbital angular momentum
by the azimuthal and polar angles ($\alpha$, $\beta$). In order to completely
specify the rotation we use the frame that minimises precessional effects
and adopt the ``minimum rotation condition'' \cite{PhysRevD.84.124011} which
enforces the third Euler angle to obey $\dot{\epsilon}(t) = \dot{\alpha}(t) \,
{\rm {cos}}(\beta(t))$.
This angle constitutes a modification to the orbital phase chosen so that the
frequency in the inertial frame is the same as the frequency in the co-precessing frame.
For a geometric depiction, see Figure~1 in
Ref.~\cite{Chatziioannou:2017tdw}\footnote{To convert from
the notation used in~\cite{Chatziioannou:2017tdw} to ours, use
the following substitutions
$\phi_z \rightarrow \alpha$, $\theta_L \rightarrow \beta$ and $\zeta \rightarrow \epsilon$.}.

This procedure was used to produce the first precessing IMR models~\cite{Hannam:2013oca,Taracchini:2013rva,PhysRevD.89.084006}.
The general procedure is as follows.
First we express the two \ac{GW} polarisations
$h_{+}$ and $h_\times$ as a linear combination
of spherical harmonics with spin-weight $-2$.
The coefficients of the basis functions are the
\ac{GW} multipoles $h_{\ell,m}$, see equation~\eqref{equ:hlm}.
This decomposition is performed in a frame that is aligned with the total
angular momentum and the direction of propagation is given by the spherical polar
coordinates ($\theta, \varphi$),

\begin{equation}
\label{equ:hlm}
h(t;\theta,\phi) = h_+ \, - \, i h_\times = \sum_{\ell \geqslant 2,m}
h_{\ell,m}(t) Y_{\ell,m}^{-2}(\theta,\varphi).
\end{equation}

From here we wish to express the \ac{GW} multipoles from precessing
\acp{BBH} $h^{prec}_{\ell,m}$
in terms of the multipoles from non-precessing \ac{BBH}s $h^{ \substack{non- \\
prec}  }_{\ell,m}$
and the appropriate angles that describe the
``twisting-up'' from non-precessional to precessional dynamics.
This is done by applying
the Wigner-D rotation matrices to the \ac{GW} multipoles from the non-precessing
system using the angles $(\alpha, \beta, \epsilon)$~\cite{PhysRevD.84.024046, PhysRevD.84.124002,PhysRevD.84.124011,PhysRevD.86.104063}.
Here we focus on the $\ell=2$ multipoles and the case where the non-precessing model only contains
the $\ell = |m| = 2$ multipoles. In that case, the waveform is given by
\begin{equation}
\label{equ:rot}
h^{prec}_{2,m}(t) = e^{-im\alpha(t)} \sum_{|m'|=2} e^{im'\epsilon(t)} \,  d^2_{m',m}(-\beta(t)) \, h^{ \substack{non- \\ prec}  }_{2,m'}(t) \, .
\end{equation}

In order to complete the tranformation from a non-precessing \ac{IMR} model
to a precessing \ac{IMR} model we also need to modify the mapping between the
inspiraling progenitor \acp{BH} and the final \ac{BH}, taking into account
the effect of precession. We typically assume that changes in the \ac{GW} flux
due to precession can be neglected so that the estimate of the final
mass can be taken from fits to \ac{NR} simulations of non-precessing systems.
The final spin, however, is sensitive to precession and therefore
needs to be modified from the model used in the underlying non-precessing model.

In the next sections, we review the angles $(\alpha_{\rm{v2}}, \beta_{\rm{v2}}, \epsilon_{\rm{v2}})$ that were used for
the {\tt IMRPhenomPv2} model and then we describe the updated $(\alpha_{\rm{v3}}, \beta_{\rm{v3}}, \epsilon_{\rm{v3}})$
that we employ to produce the updated model {\tt IMRPhenomPv3}.

\subsection{Review of {\tt IMRPhenomPv2}}

The method described in the previous section to construct
a complete \ac{IMR} model for precessing \ac{BBH}s was used to create
the {\tt IMRPhenomP} model~\cite{Hannam:2013oca}.
The original model used the {\tt IMRPhenomC}~\cite{PhysRevD.82.064016} model to describe the non-precessing system,
but was subsequently enhanced to {\tt IMRPhenomPv2}, which uses
{\tt IMRPhenomD}~\cite{PhysRevD.93.044006, Khan:2015jqa},
a more accurate aligned-spin model valid for larger
mass-ratio binaries and \acp{BH} with larger spin
magnitudes. Moreover {\tt IMRPhenomD} includes some two-spin information during the inspiral.
Both of these underlying aligned-spin models
 provide only the $\ell = |m| = 2$ spherical
harmonic multipole.

In both previous versions of {\tt IMRPhenomP} the precession angles $(\alpha_{\rm{v2}}, \beta_{\rm{v2}}, \epsilon_{\rm{v2}})$
were computed by a closed-form frequency-domain expression
derived under the assumption of a single-spin system and parameterised
by the aligned effective-spin parameter $\chi_{\rm{eff}}$ and the
precession effective-spin parameter $\chi_p$ \cite{PhysRevD.91.024043} defined
as

\begin{equation}
\label{equ:chip}
  \chi_p := \frac{\max\left( A_1 S_{1\bot}, A_2 S_{2\bot} \right)}{ A_1 m_1^2 } \, ,
\end{equation}
where $A_1 = 2+\frac{3m_2}{2m_1}$, $A_2 = 2+\frac{3m_1}{2m_2}$ and $S_{i\bot}$ are the
magnitudes of the spin-angular momenta perpendicular to $\vec{L}$.
Additionally, {\tt IMRPhenomP} used a small-angle approximation for the opening
angle ($\beta$), an assumption that was subsequently relaxed in {\tt IMRPhenomPv2}.

The expressions for $(\alpha_{\rm{v2}}, \beta_{\rm{v2}}, \epsilon_{\rm{v2}})$ were derived under the \ac{SPA}~\cite{Droz:1999qx}
and are used to ``twist-up'' the entire waveform, including
through merger and ringdown, where the assumptions of \ac{SPA} are formally
invalid. Nevertheless the extrapolation of the \ac{PN} expressions
into the merger and ringdown
have been shown to not impair the model, so long as the mass-ratio and spin are not large,
see~\cite{Hannam:2013oca} and Section~\ref{sec:mismatch}.

{\tt IMRPhenomPv2} uses a single-spin approximation that
only models the simple-precession case where the direction of $J$
is relatively constant during the inspiral, and the direction
of the final \ac{BH} spin is assumed to be parallel to $J$.
The magnitude of the final spin angular momentum $S_{f}$
is computed as the sum of parallel and perpendicular angular momentum
components with respect to $\vec{L}$. $S_{\substack{non- \\ prec}}$ is the angular momentum
from the non-precessing system and $S_{\bot}$ is the angular
momentum of the in-plane spin components.

\begin{equation}
  S_{f} \equiv M_{f}^2 \chi_{f} = \sqrt{ S^{2}_{\bot} + S^{2}_{\substack{non- \\ prec}} } \, .
\end{equation}

The final dimensionless spin magnitude $\chi_{f}$, with a free parameter $\lambda$
is written as~\cite{pv2techdoc,PhysRevD.95.064024}

\begin{equation}
  \chi_{f} = \sqrt{ \left( S_{\bot} \frac{\lambda^2}{M_{f}^2} \right)^2  + \chi^{2}_{\substack{non- \\ prec}} } \, .
\end{equation}

The choice made for $\lambda$ in {\tt IMRPhenomPv2} is $\lambda = M_f/M$
implies that the $S_{\bot}$ angular momentum gets scaled by the initial
total mass.
We comment that this is a fairly arbitrary choice as we have a model
for $M_f$ that we use in the non-precessing part~\cite{PhysRevD.93.044006}.
However, it was found that this simplification still yielded a model
with acceptable accuracy.
Finally, $S_{\bot}$ is approximated using the effective precessing parameter
$\chi_p$ and computing the angular momentum by assuming only the primary
\ac{BH} has in-plane spin components i.e.

\begin{equation}
  S_{\bot} = m_1^2 \chi_p \, .
\end{equation}

The model {\tt IMRPhenomPv2} has been used in numerous publications to estimate the source parameters
of all \ac{BBH} observations in the LIGO-Virgo O1 and O2 runs, eg.~\cite{PhysRevLett.116.131103,PhysRevLett.116.241102,PhysRevLett.116.221101,2016PhRvL.116x1103A,PhysRevX.6.041015,2017PhRvL.118v1101A,Abbott:2017oio,Abbott:2017gyy} and has been the basis of a
ROQ approximation to the likelihood~\cite{Smith:2016qas}
that has enabled fast parameter estimation of GW sources~\cite{Vitale:2017cfs}. The model's low computational cost has also enabled it to
form the basis of the model {\tt IMRPhenomPv2\_NRTidal}~\cite{Dietrich:2018uni} used for the analysis of the long BNS signal GW170817~\cite{Abbott:2018wiz,Abbott:2018exr}.

\subsection{Issues with {\tt IMRPhenomPv2} precession angles}
\label{sec:v2issues}

Despite {\tt IMRPhenomPv2}'s good performance across the parameter space, there are some known
issues with the precession angles. Specifically, the angle $\alpha_{\rm{v2}}(f)$ is written as a power series in $f$, including terms up to
next-to-next-to-leading order (3PN) in the spin. The leading-order Newtonian term (see, for example, Ref.~\cite{PhysRevD.49.6274}) is
\begin{equation}
\alpha_{\rm{v2}}^N = \alpha_0 - \frac{20\left(1 + \frac{3}{4q}\right)}{192 f}, \label{eq:aLO}
\end{equation}
where $\alpha_0$ is a constant. At this order, the precession angle is a monotonically
increasing function, which is what we expect in a physical system exhibiting simple precession: the orbital
angular momentum vector precesses around the total angular momentum throughout the binary's evolution,
and the rate of precession steadily increases. At leading order~\cite{PhysRevD.49.6274} we also see that the precession angle (and therefore
also the precession frequency)
depend only on the mass-ratio of the binary; the spin affects the precession rate at
higher orders.

If we extend this expression up to $f^{1/3}$, as is done in the
NNLO expressions used in {\tt IMRPhenomPv2}~\cite{schmidt.thesis} (which also include a $\log(f)$ term), then higher order terms can
enter with opposing signs. In some cases this can remove the monotonicity of $\alpha_{\rm{v2}}(f)$, even for simple-precession
configurations where we know that such behavior is not physically consistent.

\begin{figure*}[t]
  \includegraphics[width=0.92\columnwidth]{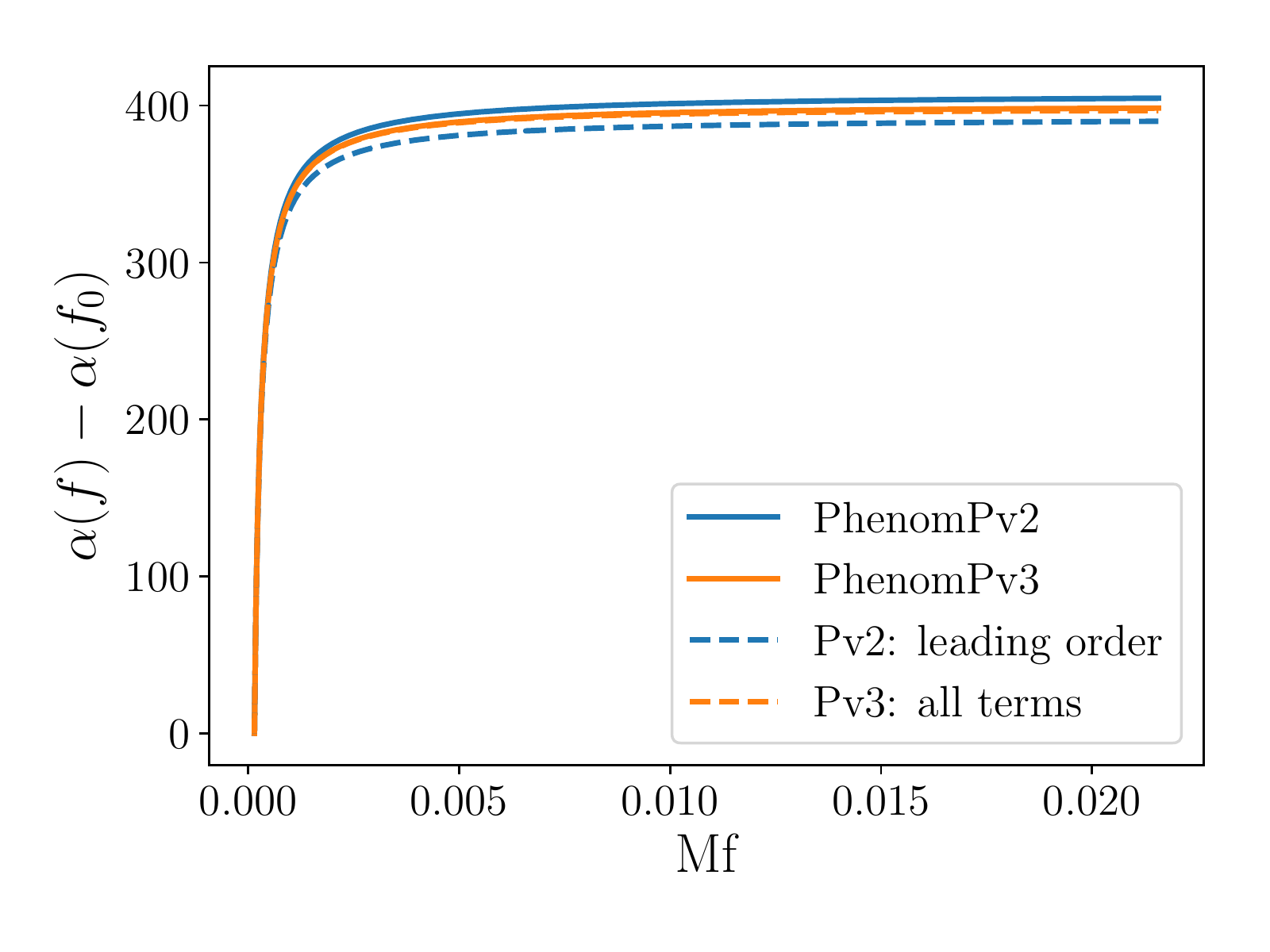}
    \includegraphics[width=0.92\columnwidth]{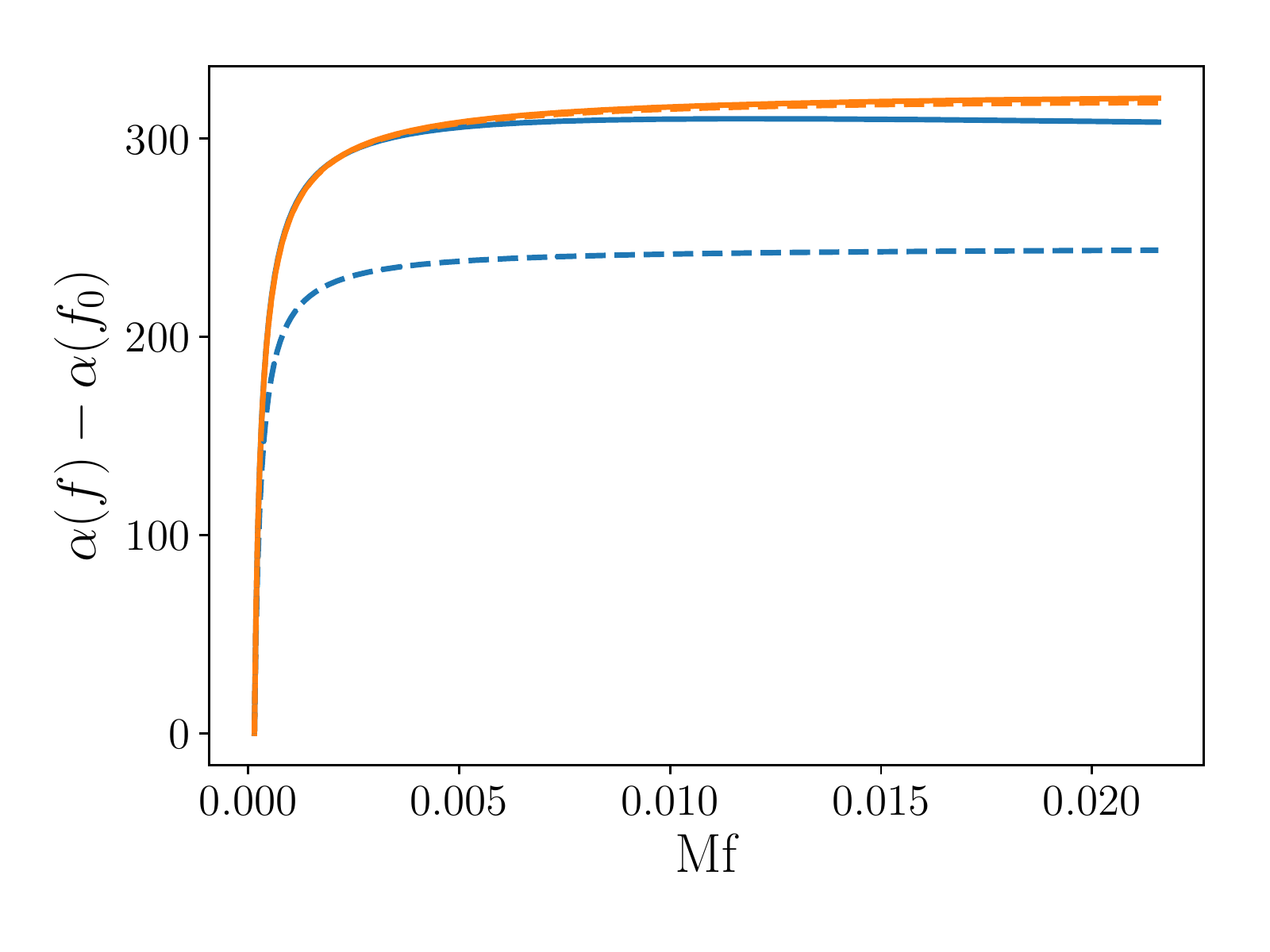}
      \includegraphics[width=0.92\columnwidth]{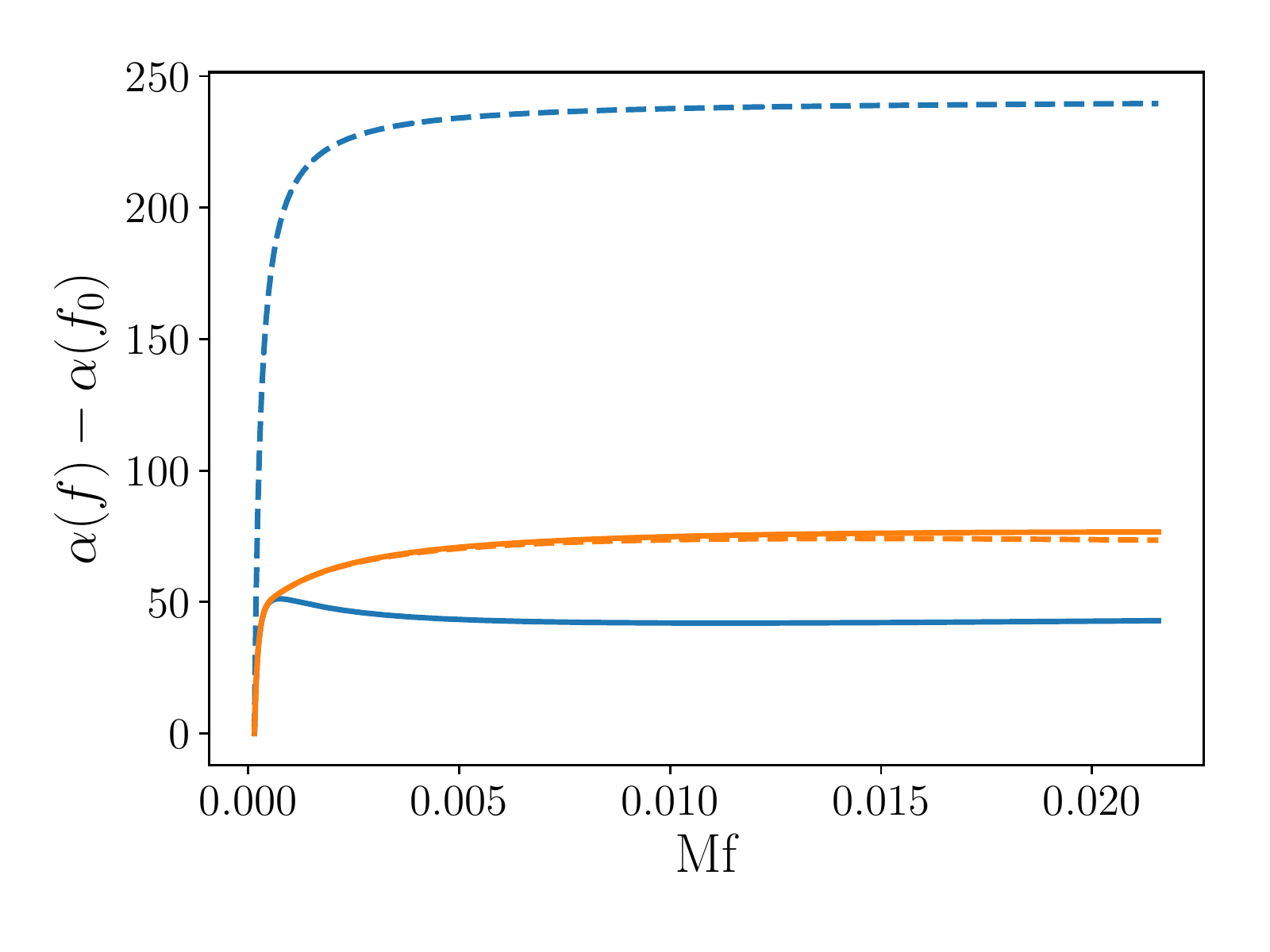}
        \includegraphics[width=0.92\columnwidth]{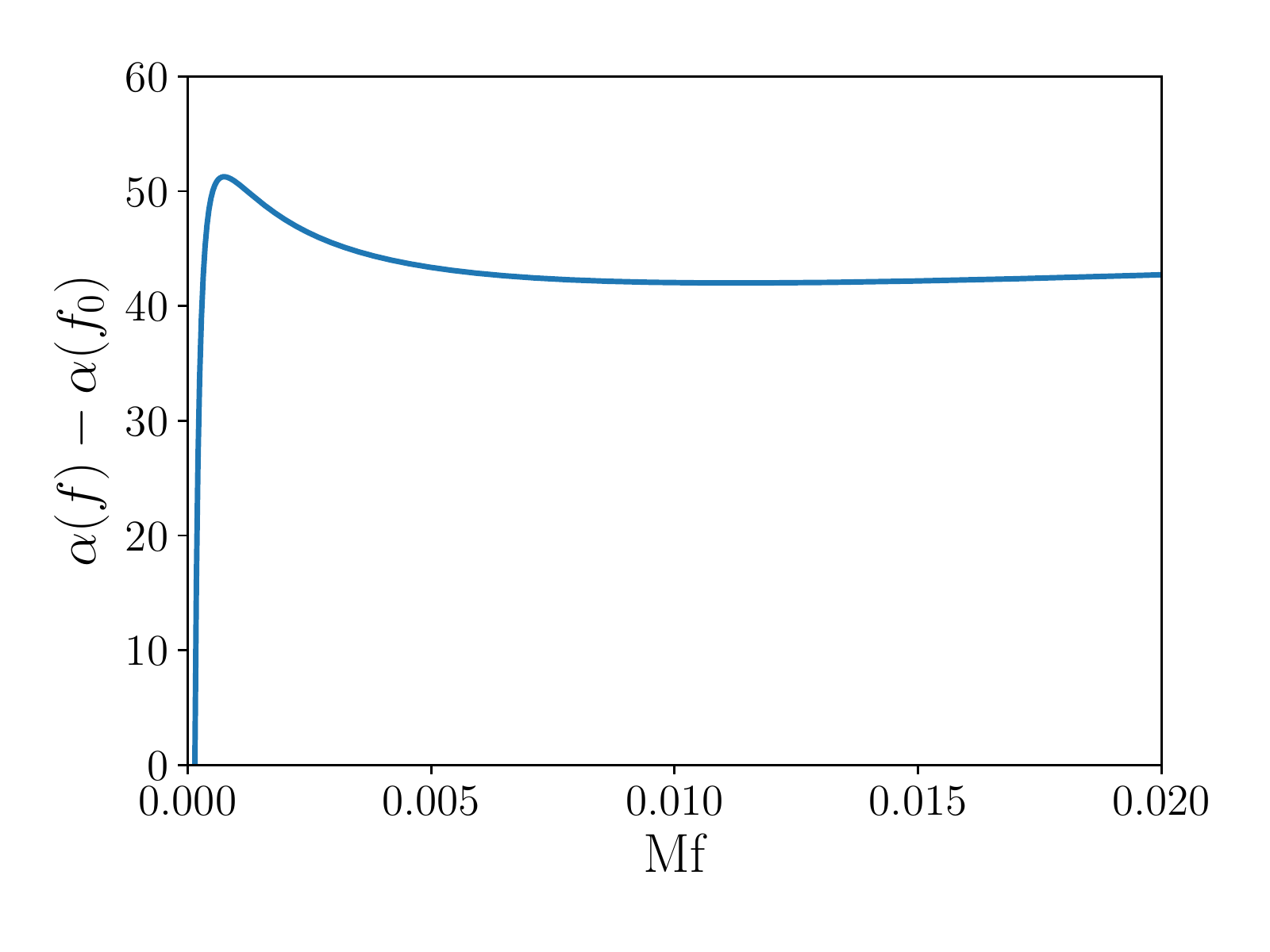}
 \caption{Comparison of leading-order ({\tt IMRPhenomPv2: leading order}), NNLO ({\tt IMRPhenomPv2}), full-PN ({\tt IMRPhenomPv3: all terms}), and
 the version we adopt in the final model ({\tt IMRPhenomPv3}) expressions for the precession angle,
 $\alpha$.
 First panel: $q=1$, $M_{{\rm{tot}}}=10 M_{\odot}$ $\chi_{\rm eff} = 0$, $\chi_p = 0.9$;
 second panel: $q=8$, $M_{{\rm{tot}}}=10 M_{\odot}$ $\chi_{\rm eff} = 0$, $\chi_p = 0.9$;
 third panel: $q=10$, $M_{{\rm{tot}}}=10 M_{\odot}$ $\chi_{\rm eff} = -0.8$, $\chi_p = 0.1$;
 fourth panel: same configuration as third panel, but zoom of non-physical behavior of $\alpha_{\rm{v2}}(f)$.
 All cases are generated from a frequency of $2$ Hz.
 See text for discussion.}
 \label{fig:AlphaComparison}
\end{figure*}

Figure~\ref{fig:AlphaComparison} show examples of this behavior. We first consider a configuration that is comparable to those already
found in GW observations: the mass ratio is 1, the component of the spin parallel to the orbital angular
momentum is zero, i.e., $\chi_{\rm eff} = 0$, and the larger BH has a spin of magnitude 0.9 lying in the orbital
plane, i.e., $\chi_p = 0.9$; since $\chi_p$ has not been constrained in observations to date, a value of $0.9$ is
consistent with the measured parameters. The angle of the spin in the plane at a particular reference frequency determines
the constant $\alpha_0$, but we will focus here on the frequency evolution of the precession angle. We plot
the precession angle $\alpha_{\rm{v2}}(f) - \alpha(f_0)$, where $M f_0 = 0.0001$, which corresponds to approximately 2\,Hz for a
binary with a total mass of 10\,$M_\odot$. The results are shown up to the
Schwarzschild ISCO frequency ($Mf = 0.0217$).
We can see that $\alpha_{\rm{v2}}(f)$, as predicted by the leading-order Eq.~\ref{eq:aLO} (blue dashed line),
the full NNLO expressions used for {\tt IMRPhenomPv2} (blue solid line), the full {\tt IMRPhenomPv3} expression (orange dashed line), and the trunchated {\tt IMRPhenomPv3} expression (orange solid line)
all agree well, even though spin effects are not included in the leading-order result.

In the second example, we can see that the NNLO expression is needed to accurately describe the precession for higher
mass ratios and higher spins. This example shows $\alpha_{\rm{v2}}(f) - \alpha(f_0)$ for a binary with mass-ratio 8, again $\chi_{\rm eff} = 0$ and $\chi_p = 0.9$.
The NNLO {\tt IMRPhenomPv2} and the {\tt IMRPhenomPv3} expressions agree well, but the
leading-order expression accumulates a difference against the others of $\sim$60\,rad, or $\sim$10 precession cycles, out of $\sim$47
cycles over the course of the entire inspiral. This level of agreement between the {\tt IMRPhenomPv2} and {\tt IMRPhenomPv3} expressions
is typical across all configurations with mass ratios up to $\sim$5,
so we can be confident that {\tt IMRPhenomPv2} accurately models the
precession angle $\alpha_{\rm{v2}}$ with sufficient accuracy for aLIGO and Virgo BBH observations to date.

In the third example, the mass-ratio is 10, but now $\chi_{\rm eff} = -0.8$, $\chi_p = 0.1$.
Now the {\tt IMRPhenomPv2} expressions disagree significantly between each other
and against the {\tt IMRPhenomPv3} expressinos.
Most notable, however, is that the {\tt IMRPhenomPv2} precession angle (shown alone in the fourth panel of Fig.~\ref{fig:AlphaComparison})
reaches a maximum and then decreases; the
maximum implies that at this point the precession comes to a halt, and then continues in the opposite direction. This occurs at
a frequency of $\sim10$\,Hz for a 10\,$M_\odot$ binary, i.e., many orbits before merger, while the system should still be undergoing
simple precession, and when PN results should still be valid. Indeed, simple precession does continue through these frequencies in an
evolution of the PN equations of motion, as indicated by the $\alpha_{\rm{v3}}(f)$ results shown in the third panel.
We therefore conclude that the NNLO frequency expansion behaves unphysically for certain
mass-ratio and spin combinations, and will degrade the accuracy of a model in these regions of parameter space.

\subsection{Upgrading to  {\tt IMRPhenomPv3}}

The first closed-form analytic inspiral waveform model for generically precessing BBHs was  presented in \cite{Chatziioannou:2017tdw,Chatziioannou:2016ezg}.
The solution utilized two ingredients.
First, the analytic solution to the conservative precession
equations constructed in \cite{PhysRevLett.114.081103} was supplemented by radiation-reaction effects through a
perturbative expansion in the ratio
of the precession to the radiation reaction time scale~\cite{Klein:2013qda,Chatziioannou:2013dza} known as \ac{MSA}~\cite{Bender}.
Second, the frequency-domain waveform was analytically computed through the method of \ac{SUA}, first introduced in
\cite{PhysRevD.90.124029}. The resulting inspiral waveform model for the precession dynamics incorporates spin-orbit and spin-spin
effects to leading order in the conservative dynamics and up to
3.5PN order in the dissipative dynamics ignoring spin-spin terms.

We use the generic two-spin solution of \cite{Chatziioannou:2017tdw,Chatziioannou:2016ezg} to obtain two-spin expressions for the precession angles
 $(\alpha_{\rm{v3}}, \beta_{\rm{v3}}, \epsilon_{\rm{v3}})$.
Specifically we use Eqs. (58), (66), (67) and the coefficients in Appendix D in~\cite{Chatziioannou:2017tdw} for $\alpha_{\rm{v3}}$ (denoted as $\phi_z$ in that paper). Appendix F in~\cite{Chatziioannou:2017tdw} provides the analogous equations
and coefficients for the $\epsilon_{\rm{v3}}$ angle (called $\zeta$ in~\cite{Chatziioannou:2017tdw}), while the $\beta_{\rm{v3}}$ angle (called $\theta_{L}$ in~\cite{Chatziioannou:2017tdw}) is given by Equation (8).

\begin{figure*}[t]
  \includegraphics[width=\textwidth]{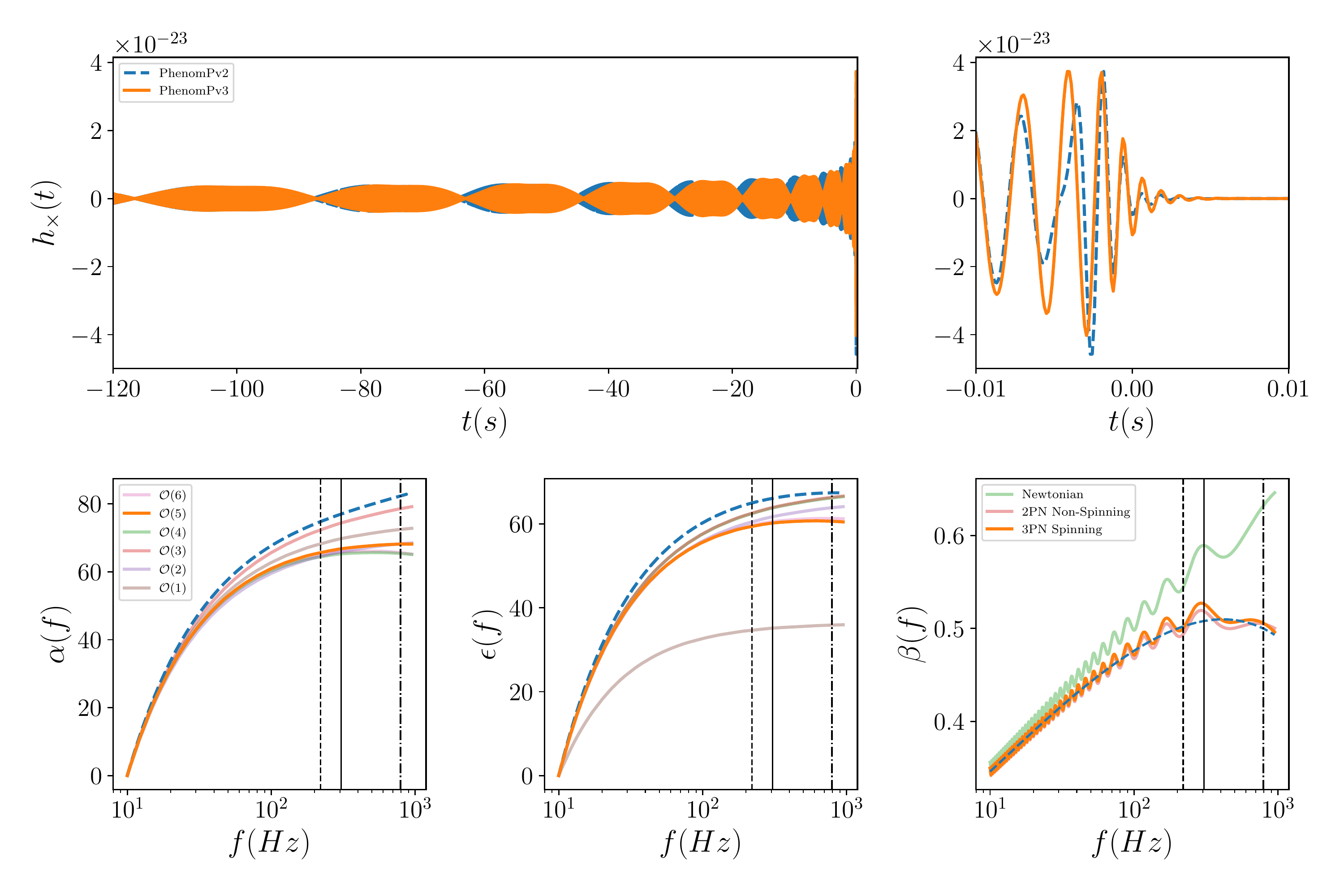}
 \caption{
 Composite figure comparing the various models for the precession
 angles and their overall effect on the waveform.
 We generate the gravitational waveform from a \ac{BBH} system with the
 following parameter:
 $q=10$, $M=20$, $S_1=(0.4,0,0.4)$, $S_2=(0.3,0,-0.3)$, $f_{{\rm start}} = 10$Hz.
 The top row shows $h_{\times}$ polarisation evaluated at
 $\iota = \pi/2$ (Here $\iota$ is the angle between $\vec{L}$ and the
 line of sight).
 In all panels the dashed blue line corresponds to {\tt IMRPhenomPv2} and the
 solid orange line corresponds to {\tt IMRPhenomPv3}.
 Top left: The full waveform generated from 10Hz.
 Top right: Zoom in around merger.
 The early-time agreement and late-time disagreement demonstrates
 the agreement between the two models of precession at low frequencies
 and disagreement at high frequencies.
 Bottom row shows the precession angles as a function of \ac{GW} frequency.
 Bottom left: $\alpha$. Bottom middle: $\epsilon$. Bottom right: $\beta$.
 The solid black line is the hybrid-MECO
 frequency \cite{PhysRevD.95.064016},
 the dashed black line
 is the Schwarzschild ISCO frequency and the dot-dashed black line is the ringdown
 frequency for this system.
 The legend in the bottom left plot applies to the bottom middle plot
 as well and shows to what $n$ order (see text) the $\alpha$
 and $\epsilon$ angles where expanded.
 The legend in the bottom right figure indicates if aligned-spins
 were included in the calculation of the magnitude of the orbital angular momentum.
 }
 \label{fig:waveform-and-angle}
\end{figure*}

As an illustration of the new waveform model we compare {\tt IMRPhenomPv2} (blue-dashed)
and {\tt IMRPhenomPv3} (orange-solid) for a mass-ratio 1:10, two-spin system
in Figure~\ref{fig:waveform-and-angle}.
The top left panel shows the
$h_{\times}(t)$\footnote{The time domain $h_{\times}(t)$ was obtained by computing the inverse Fourier
transform of $\tilde{h}_{\times}(f)$.}
\ac{GW} polarisation
viewed at an inclination angle\footnote{Inclination is the angle between the $\vec{L}$
and the line of sight.}
of $90 \, \rm{deg}$. The top right panel is a zoom in around the merger.
At early times both models are in agreement however, due to the difference
between the precession angle models the two models start to noticeably disagree
around $90 \, \rm{s}$ before merger.

Similarly to the angles $\alpha_{\rm{v2}}$ and $\epsilon_{\rm{v2}}$, the expressions for $\alpha_{\rm{v3}}$ and $\epsilon_{\rm{v3}}$ involve
series expansions in terms of the GW frequency. The expression for $\epsilon_{\rm{v3}}$ is fully expanded and expressed in terms of a power series of the
\ac{GW} frequency, $f^{-4/3+n/3}$, $n \in [1,6]$. The angle $\alpha_{\rm{v3}}$ involves both expanded ($n \in [1,6]$) and un-expanded terms,
a choice made to increase the angle's accuracy for unequal-mass systems, as discussed in Sec. IV D 1 of~\cite{Chatziioannou:2017tdw}.
The order to which we truncate the relevant expansions
can impact the accuracy of the model: too few $n$ terms and the expansion
fails to accurately describe the inspiral but too many $n$ terms and the
expressions become inaccurate when extrapolating towards higher frequencies.
Figure~\ref{fig:waveform-and-angle} illustrates the impact of expansion
order $n$ on the precession angles.
The lower-left and lower-middle panels show the $\alpha_{\rm{v3}}$ and $\epsilon_{\rm{v3}}$ angles
as a function of the \ac{GW} frequency.
The solid-orange line shows the model that is
used in {\tt IMRPhenomPv3} and the paler curves show different the result
for truncation orders for the $\alpha_{\rm{v3}}$ and $\epsilon_{\rm{v3}}$ models.
The dashed-blue line shows the results for the precession angle model
used in {\tt IMRPhenomPv2} which deviates away from the other
approximations. The $\alpha_{\rm{v2}}$ angle in particular shows qualitatively different
behaviour, growing rapidly as the frequency increases.

We also have a choice when computing $\beta_{\rm{v3}}$, the angle between
$\vec{J}$ and $\vec{L}$.
We investigated either using Newtonian order, 2PN Non-Spinning
(as was done in {\tt IMRPhenomPv2}) and also a 3PN version including
spin-orbit terms for the magnitude of $\vec{L}$ that is used in the computation
of $\beta_{\rm{v3}}$.
A comparison between the different methods for calculating $\beta_{\rm{v3}}$
can be seen in the lower-right panel in Fig.~\ref{fig:waveform-and-angle}.
The observed modulations are nutation due to spin-spin effects.

The three vertical black lines (from left to right) are the
hybrid-MECO~\cite{PhysRevD.95.064016},
Schwarzschild ISCO and the ringdown frequency for this system.
We assume the limit to which \ac{PN} results can be reliably used to be near
the hybrid-MECO/Schwarzschild ISCO and the region between this and the
ringdown frequency to be the region where we are extrapolating the
precession angles beyond their assumed region of validity.
This assumption appears to hold as it seems to track the location
of a turn-over point in many of the precession angle variants, a feature that
is unphysical (at least in simple precession cases) as discussed in
Sec.~\ref{sec:v2issues}.

To decide which choices when computing the precession anlges
lead to the most accurate model we performed several mismatch calculation
comparing different versions of the model with \ac{NR}.
We find that most choices lead to reasonably accurate models for the inspiral
but, however, can lead to suboptimal performance during the merger.
We obtain an accurate model for the entire coalescence by using
all but the highest order terms in the expressions for $\alpha_{\rm{v3}}$ and $\epsilon_{\rm{v3}}$
and to use the highest order PN calculation available namely, with
3PN with spin-orbit terms when computing $|\vec{L}|$ in the $\beta$ angle.
In Sec.~\ref{sec:mismatch} we only show results for the final model.

To minimise differences bewteen {\tt IMRPhenomPv2} and {\tt IMRPhenomPv3}
we adopt the same model for the final spin however, we note
a simple extension would be to include all in-plane spin components
in the calculation of $S_{\bot}$ and relaxing the approximation
that $M_{f}=M_{i}$ by using the final mass model used
in the non-precessing case. This will be done in following work.

We also note one other improvement on {\tt IMRPhenomPv2}, which deals with ``superkick''
configurations~\cite{Brugmann:2007zj}. These are equal-mass configurations where each black
hole has the same spin, but the spins both lie in the orbital plane and point in opposite directions.
These configurations possess symmetry such that the orbital plane does not precess, but instead
``bobs'' up and down as linear momentum is radiated perpendicular to the orbital plane. If
{\tt IMRPhenomPv2} is given the parameters for such a configuration, the definition of $\chi_p$,
Eq.~(\ref{equ:chip}) will yield a non-zero value, and the code will construct the waveform for a precessing system.
Neighbouring configurations in parameter space will display similar behaviour, and {\tt IMRPhenomPv2} will
again generate the ``wrong'' waveform. By correctly accounting for both spins, {\tt IMRPhenomPv3}
corrects this problem. A request for an equal-mass superkick configuration will yield the waveform
for an equal-mass nonspinning configuration, which is the closest approximation that does not
experience recoil.

\section{Comparison to numerical relativity}
\label{sec:mismatch}

In this section we compare {\tt IMRPhenomPv3} to a large number of \ac{NR} simulations of precessing \ac{BBH}s and show the excellent
agreement between our model and the simulations.

\subsection{Mismatch calculation}

In GW searches and parameter estimation, template
waveforms are correlated with detector data.
This operation can be written as an inner-product weighted by
the sensitivity of the detector (described by the \ac{PSD} $S_n(f)$) between the real valued template $h(t)$ and signal $s(t)$
waveforms.
We define the \emph{overlap} between the template and signal as

\begin{equation}
\label{equ:overlap}
\mathcal{O} \equiv \left(  h | s  \right) \equiv \, 4 \, {\rm{Re}} \int\displaylimits_{f_{\rm{min}}}^{f_{\rm{max}}}  \frac{ \tilde{h}^*(f) \tilde{s}(f)   }{S_n(f)} df \, ,
\end{equation}
where $\tilde{h}$ representes the Fourier transform of $h$ and
$\tilde{h}^*$ is the complex conjugate of $\tilde{h}$.

This inner-product is closely related to the definition of the matched-filter
\ac{SNR} and indeed the $1 - \mathcal{O}$ between normalised waveforms
($\hat{x} = \sqrt{ (x|x) }$)
is directly proportional to the loss in \ac{SNR} making it a useful metric
to measure the accuracy of waveform models.

Aside from the masses and spins of the BHs the GW
signal depends on a number of extrinsic parameters: the direction
of propagation in the source frame ($\iota, \phi_0$),
a polarisation angle $\psi$,
a reference time $t_0$
and the luminosity distance $D_{L}$.
The dependency of the observed \ac{GW} signal on the polarisation angle
(ignoring the angular position of the source with respect to a detector)
is given by

\begin{equation}
\label{equ:hresp}
h(t) = h_+(t) \, {\rm{cos}}(2\psi) + h_\times(t) \, {\rm{sin}}(2\psi) \, .
\end{equation}
Given two waveforms $h$ and $s$ we quantify the agreement between them
by computing their normalised overlap subject to various averages and
optimisations of extrinsic parameters whilst keeping
the intrinsic parameters fixed.
Note that the use of normalised waveforms $\hat{h}$ and $\hat{s}$ factors away the dependency
of the $D_{L}$ in the overlap.

In particular for each inclination angle $\iota$ considered we
compute the overlap between $\hat{h}$ and $\hat{s}$ with the same
masses and spins and vary the extrinsic \emph{signal} parameters
($\phi^s_0, \psi^s_0$) by evaluating them on
a $10 \times 10$ grid with the following domain
$\phi^s_0 \in [0, 2\pi]$ and $\psi^s_0 \in [0, \pi/4]$.
For each point on this grid we analytically maximise over a
time shift using an inverse Fourier transform, analytically
maximise over $\psi^h_0$ according to the method detailed in
\cite{PhysRevD.94.024012} and finally numerically optimise over $\phi^h_0$
using optimisation routines from the {\tt SciPy} package
\footnote{Because we are comparing generic precessing waveforms
we can not analytically optimise over $\phi^h_0$ in the presence of
$m \neq 2$ multipoles.}.
From here we define the \emph{match} as a function of the extrinsic
parameters of the signal as

\begin{equation}
\mathcal{M}(\phi^s_0, \psi^s_0) \equiv \max_{t^h_0, \phi^h_0, \psi^h_0} \, (  \hat{h} | \hat{s}(\phi^s_0, \psi^s_0) ) \, .
\end{equation}
Next we average over the extrinsic parameters of the signal $(\phi^s_0, \psi^s_0)$
weighted by
its optimal \ac{SNR} $\rho$ at each point which accounts for the
likelihood that this signal would be detected~\cite{Buonanno:2002fy,Harry:2017weg}.
We call this the \emph{orientation-averaged-match}

\begin{equation}
\overline{\mathcal{M}} \equiv \left(  \frac{\sum\limits_{i} \rho_i^3 \mathcal{M}_i^3}{\sum\limits_{i} \rho_i^3} \right)^{1/3} \, .
\end{equation}
From here we define the \emph{orientation-averaged-mismatch} as
$1 - \overline{\mathcal{M}}$. We will quote results in terms of this.
In what follows we shall use $h$ to denote a \emph{template} waveform, i.e.,
generated by a waveform approximant and $s$ to denote the \emph{signal}
waveform which will be an \ac{NR} waveform.

We compute the orientation-averaged-mismatch between the template waveform
and each \ac{NR} waveform at three different inclination angles
$(0, \pi/3, \pi/2) \, \rm{rad}$.
By increasing the inclination angle we tend to observe a weaker but
more modulated waveform due to the precession.
We therefore expect the accuracy of the models to decrease at
inclined orientations where the effects of precession are typically
more prounounced.
We consider 3 different waveform approximants: {\tt IMRPhenomPv2},
our improvement {\tt IMRPhenomPv3} and the precessing \ac{IMR} EOB-NR
model {\tt SEOBNRv3}~\cite{PhysRevD.95.024010}.
For each \ac{NR} simulation we generate a template
with the same masses and spins beginning from the start frequency
as reported in the \ac{NR} metadata.
Waveforms were generated using the {\tt LALSimulation} package, part of the
software library {\tt LALSuite} \cite{lalsuite}, using the \ac{NR}
injection infrastructure presented in~\cite{Schmidt:2017btt}.

When comparing \ac{GW} signal models with \ac{NR}
there is an ambiguity that one encounters when trying to identify a
time (or frequency) in the gravitational waveform and the corresponding
retarded time (or frequency) of the \ac{BBH} dynamics, where spins are
measured and defined in the \ac{NR} simulation~\cite{Hamilton:2018fxk}.
This complicates the comparison of precessing systems because the orientation
of the spin is now time dependent.

To account for the ambiguity in how the spins are defined we allow
the frequency at which the spins are specified to vary.
This is similar to what was done in Ref.~\cite{PhysRevD.95.024010}
where {\tt SEOBNRv3} was compared to a similar set of \ac{NR} waveforms.
In {\tt IMRPhenomPv2} and {\tt IMRPhenomPv3} we can fix the start frequency, $f_{\rm{start}}$,
and vary the spin-reference-frequency ($f_{\rm{ref}}$), however,
in the {\tt LALSuite} implementation of {\tt SEOBNRv3} this is not possible
and instead the spins are defined at the start frequency.
Therefore we perform the optimisation of spin-reference-frequency
slightly differently between the two phenom models and the EOB-NR model.
We numerically optimise using {\tt SciPy} for both cases.
For the phenom models we allow $f_{\rm{ref}}$ to vary
in the following range $[0.8, 1.4] \times f_{\rm{start}}$
and for {\tt SEOBNRv3} we vary $f_{\rm{start}}$
in the same range but if $1.4 f_{\rm{start}}$ is greater than the
maximum start frequency allowed by the EOB-NR generator, $f^{\rm{EOB}}_{\rm{max}}$,
then we use this\footnote{$f^{\rm{EOB}}_{\rm{max}}$ is equal to the
the orbital frequency at a separation of $10M$.}.
For long waveforms the variation in the resulting match is less that 1\%
however, for shorter waveforms, such as SXS:BBH:0165, the variation can be as larger as 8\%.

Our \ac{NR} signal waveforms only contain the $\ell = 2$ multipoles
which is a good approximation for comparable mass systems $q \lesssim 3$
and for small inclinations where the effect of precession on the waveform
is minimised. This assumption breaks down for some of the cases we consider
here but we are primarily concerned with how faithful our new model is to a
signal which contains the modes that we have modelled. We also want to keep
separate systematic errors due to neglecting $\ell>2$ modes and those due to
inaccuracies in modelling the $\ell=2$ modes.
We compute the match assuming the theoretical design sensitivity of one of the LIGO detectors
(zero de-tuned high power \ac{PSD}~\cite{aligozetodethp}) and starting from a low
frequency cutoff of $10 \, \text{Hz}$ until the end of the waveform. We compute the match
over a range of total masses spanning from $20 \, M_\odot$ to $200 \, M_\odot$.
If the NR waveform is not long enough to start at $10 \, \text{Hz}$ then
we use the starting frequency of the NR waveform as the low frequency cutoff in
the match integral.

\subsection{NR catalogue}

\begin{figure*}[t]
  \includegraphics[width=\textwidth]{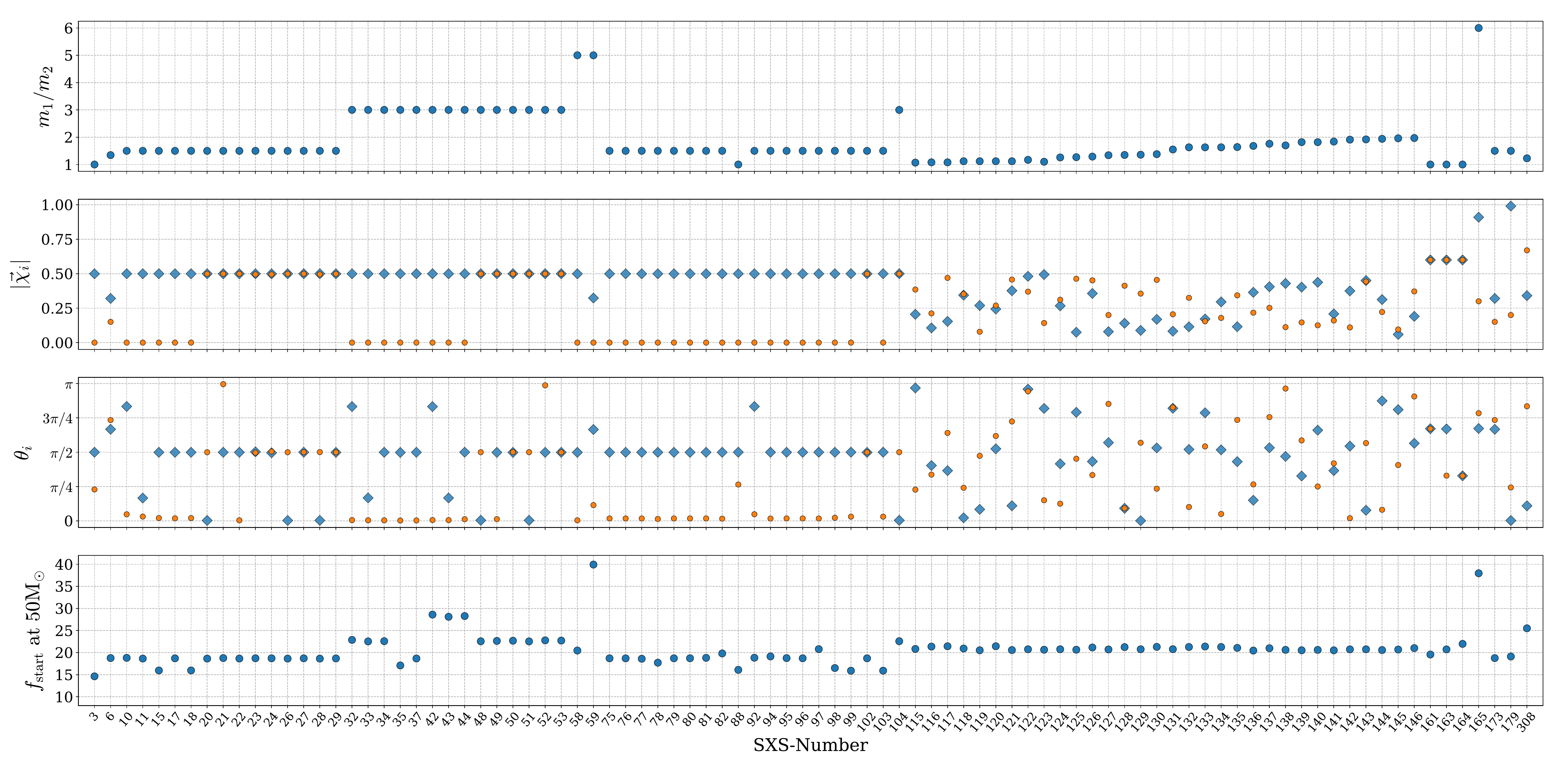}
 \caption{The precessing parameter space we probe using the SXS public
 catalogue. The x-axis is the SXS catalogue number. The four panels
 (from top to bottom) show the mass-ratio $q = m_1/m_2$, the dimensionless
 spin magnitudes $|\vec{\chi}_i|$, the polar angle between the Newtonian orbital
 angular momentum and the individal BH spin vectors $\theta_i$ where $i=1,2$
 for the larger and smaller BH respectively and finally the last panel
 plots the start GW frequency for each simulation when scaled to a total
 mass of $50 \rm{M_{\odot}}$.
 Where the primary and secondary BHs are plotted as blue diamonds and
 orange circles respectively.}
 \label{fig:parspace}
\end{figure*}

We use the SXS public catalogue \cite{PhysRevLett.111.241104,sxsonline} of \ac{NR} waveforms
as our validation set consisting of 90 precessing
waveforms with mass-ratios between 1:1 and 1:6. Figure~\ref{fig:parspace}
presents an illustration of the parameter space covered.
This set of \ac{NR} waveforms mainly covers the mass-ratio space between 1:1
and 1:3 with only 3 waveforms above this: 2 at mass-ratio 1:5 and one
at 1:6. This is clearly a heavily undersampled region of parameter space
across \ac{NR} groups (see other public catalogues from
RIT \cite{0264-9381-34-22-224001, ritonline} and
GaTech \cite{0264-9381-33-20-204001, gatechonline}).
The vast majority of cases also have spin magnitudes that are $\leqslant 0.5$
however there is one case (SXS:BBH:0165) where $|\vec{\chi}| = 0.9$. However,
this is a short waveform of only 6 orbits in length. Again this highlights
the need for longer \ac{NR} simulations of \acp{BBH} with $|\vec{\chi}| > 0.5$
across all mass-ratios.

The bottom panel of Fig.~\ref{fig:parspace} illustrates the length of
the \ac{NR} simulations. It shows the start frequency, in Hz, of the
$\ell = |m| = 2$ \ac{GW} multipole when the \ac{BBH} system is scaled
to $50 M_\odot$.
In order for a \ac{NR} waveform to be used in the analysis of LIGO-Virgo
data, without hybridising to \ac{PN}, the waveform needs to
span the sensitive region of the detector, a typical start frequency
for current instruments is between $20-25 \rm{Hz}$ (see Figure 2 of~\cite{Abbott:2017oio}).
We see that for a \ac{BBH} of total mass $50 M_\odot$, similar to GW150914~\cite{2016PhRvL.116f1102A},
that many of these cases start at around $20 \rm{Hz}$.
Again there are a couple of outliers at mass-ratios 1:5 and 1:6.

\subsection{Results}

\begin{figure*}[t]
  \includegraphics[width=\textwidth]{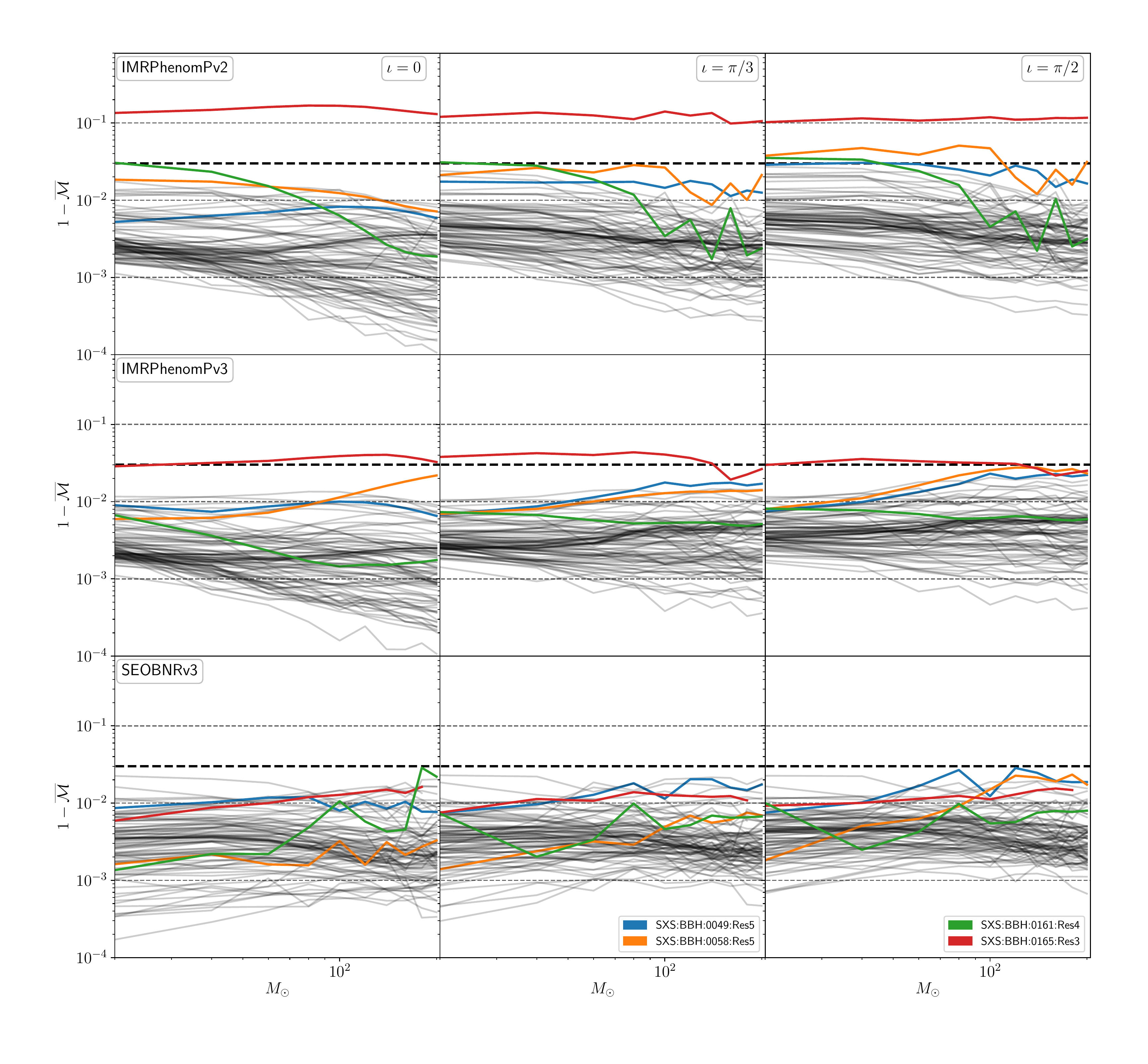}
 \caption{
   The results of the comparison between {\tt IMRPhenomPv2} (first row),
 {\tt IMRPhenomPv3} (second row) and {\tt SEOBNRv3} (third row)
 and the precessing \ac{NR} simulations from the public SXS catalogue.
 Each column shows the results for $\iota = (0, \pi/3, \pi/2)$,
 from left to right. Where $\iota$ is the angle between the
 Newtonian orbital angular momentum $\vec{L}$ and the line-of-sight,
 at the start frequency.
 The figure shows the orientation-averaged-mismatch
 ($1 - \overline{\mathcal{M}}$) as a function of the total mass (log scale).
 Cases which have a maximum mismatch greater than $3\%$ are coloured
 and presented in Table~\ref{table:worst-cases}.
 }
 \label{fig:matches}
\end{figure*}

Figure~\ref{fig:matches}  shows the results of the
orientation-averaged-mismatch ($1 - \overline{\mathcal{M}}$)
calculation as a function of the total mass.
Each row corresponds to a fixed template waveform
labeled in the top left of each plot in the first column.
From top to bottom each row show results for {\tt IMRPhenomPv2},
{\tt IMRPhenomPv3} and {\tt SEOBNRv3}.
From left to right each column corresponds to the three different
inclination angles we tested $0, \pi/3$ and $\pi/2$ respectively.

First we note that, for all three models, the majority of cases have
mismatches smaller that $3\%$ and many smaller than $1\%$.
This is in agreement with the findings of Ref~\cite{PhysRevD.95.024010}.
Interestingly we find that the mismatch varies weakly with inclination angle
indicating that the models perform well even when precession effects are
amplified.
One of the key results is the superior performance of {\tt IMRPhenomPv3}
for total masses between
$20-50 M_\odot$ where all but one case (SXS:BBH:0165 is discussed below)
have mismatches $\lesssim 1\%$.
We attribute this to the more accurate and reliable new model
of the precession dynamics of~\cite{Chatziioannou:2017tdw,Chatziioannou:2016ezg}.

To highlight regions of parameter space where current models are
least accurate we have compiled Table~\ref{table:worst-cases},
which lists the SXS catalogue number, their mass-ratio
and initial spin components if at \emph{any} point the
mismatch is greater than $3\%$.
In the table we report the orientation-averaged-mismatch averaged over all total
masses for each inclination angle.
We also colour these cases in Figure~\ref{fig:matches}.

Our expectation that higher mass-ratio systems will yield the least accurate
results due to the fact that precession is based on a \ac{PN} model
and not calibrated to \ac{NR}, is borne out in our results.
Out of all the worst cases, with the exception of SXS:BBH:0161, the worst cases
have mass-ratios $\geqslant 3$.
SXS:BBH:0161 is the only equal-mass outlier with worst mismatch marginally
greater than $3\, \%$ for {\tt IMRPhenomPv2}. For {\tt IMRPhenomPv3}
and {\tt SEOBNRv3} the accuracy is better than $1\, \%$ for all the vast
majority of cases.

\begin{table*}[t]
\centering
\begin{tabular}{l|lll|lll|lll}
  \hline \hline
Waveform Model                    & \multicolumn{3}{l|}{{\tt IMRPhenomPv2}} & \multicolumn{3}{l|}{{\tt IMRPhenomPv3}} & \multicolumn{3}{l}{{\tt SEOBNRv3}} \\ \hline
SXS:BBH:ID                          & $0$           & $\pi/3$       & $\pi/2$      & $0$     & $\pi/3$     & $\pi/2$    & $0$       & $\pi/3$      & $\pi/2$     \\ \hline
\multicolumn{1}{l|}{0049 [$q = 3$, $\vec{S}_1 = (0.5, 0, 0)$, $\vec{S}_2 = (0, 0, 0.5)$]}    & 0.7           & 1.5           & 2.4          & 0.9     & 1.4         & 1.8        & 1.0       & 1.5          & 1.8         \\
\multicolumn{1}{l|}{0058 [$q = 5$, $\vec{S}_1 = (0.5, 0, 0)$, $\vec{S}_2 = (0, 0, 0)$]}    & 1.2           & 2.0           & 3.3          & 1.3     & 1.2         & 2.2        & 0.2       & 0.5          & 1.4         \\
\multicolumn{1}{l|}{0161 [$q = 1$, $\vec{S}_1 = (0.52, 0, -0.3)$, $\vec{S}_2 = (0.52, 0, -0.3)$]}    & 1.0           & 1.1           & 1.4          & 0.2     & 0.6         & 0.7        & 0.8       & 0.6          & 0.7         \\
\multicolumn{1}{l|}{0165 [$q = 6$, $\vec{S}_1 = (0.74, 0.19, -0.50)$, $\vec{S}_2 = (-0.19, 0, -0.23)$]}    & 15.4          & 12.5          & 12.1         & 3.6     & 3.5         & 3.0        & 1.2       & 1.1          & 1.3         \\ \hline \hline
\end{tabular}
\caption{
  Percentage mismatches averaged over all total masses for each the
three inclination angles (rad) considered. All cases that have a mismatch larger
than $3\%$ are shown.
The first column shows the SXS ID number with the mass-ratio and initial spin vectors in parentheses.}
\label{table:worst-cases}
\end{table*}

The case with the highest mismatch, across all templates, is SXS:BBH:0165.
This is a mass-ratio 1:6 system with initial spins $S_1 = (0.74, 0.19, -0.50)$
and $S_2 = (-0.19, 0, -0.23)$ with a length of $\sim 6.5$ orbits.
{\tt IMRPhenomPv2} has a best mismatch of $\sim 12\%$.
{\tt IMRPhenomPv3} substantionally improves upon this with a worst
mismatch of $\sim 3.6\%$. {\tt SEOBNRv3} performs well achieving
a worst mismatch of $\sim 1.3\%$.
Given that the precession, in all three \ac{IMR} models,
is modelled with uncalibrated \ac{PN} or \ac{EOB} calculations, it is not surprising
that the region of parameter space where the models do worst comes from
high mass-ratios and high spin magnitudes.
This is, by far, the case with the most dramatic improvement.
In terms of its place in parameter space it has the largest
mass-ratio and spin magnitude.

Out of the two mass-ratio 1:5 systems one of them has a
worst mismatch larger than $3\%$. This case, SXS:BBH:0058, is also the longest
mass-ratio 1:5 case with approximately 28 orbits.
We improve from a worst mismatch of $3.3\%$ with {\tt IMRPhenomPv2}
to $2.2\%$ with {\tt IMRPhenomPv3}. {\tt SEOBNRv3}
shows good agreement with a worst average mismatch of $\sim 1.4\,\%$.

For cases with mass-ratio 1:3 we find that only 1 (SXS:BBH:0049) out of the 15 cases
has a mismatch larger than $3\%$.
The mismatch is worst for {\tt IMRPhenomPv2} and only goes marginally above $3\,\%$
for the $\iota = \pi/2$ case where we expect the effects of precession to
be most pronouced.
With {\tt IMRPhenomPv3} we improve, across all inclinations, with
average mismatches between $1-2\,\%$. We find a similar performance
with {\tt SEOBNRv3}.

Note however, that all the mass-ratio 1:3 and 1:5
systems all have spin magnitudes $\leqslant 0.5$.
Validating precessing waveform models at large spin magnitudes requires
more \ac{NR} simulations to be performed.
We highlight again that these comparisons only include the $\ell = 2$
multipoles in the signal waveform and for larger mass-ratio systems
$(q \gtrsim 3)$ negleting the higher multipoles is no longer a good
approximation to the full \ac{GW} signal.

\section{GW151226 analysis}

\begin{figure*}[t]
  \includegraphics[width=\columnwidth]{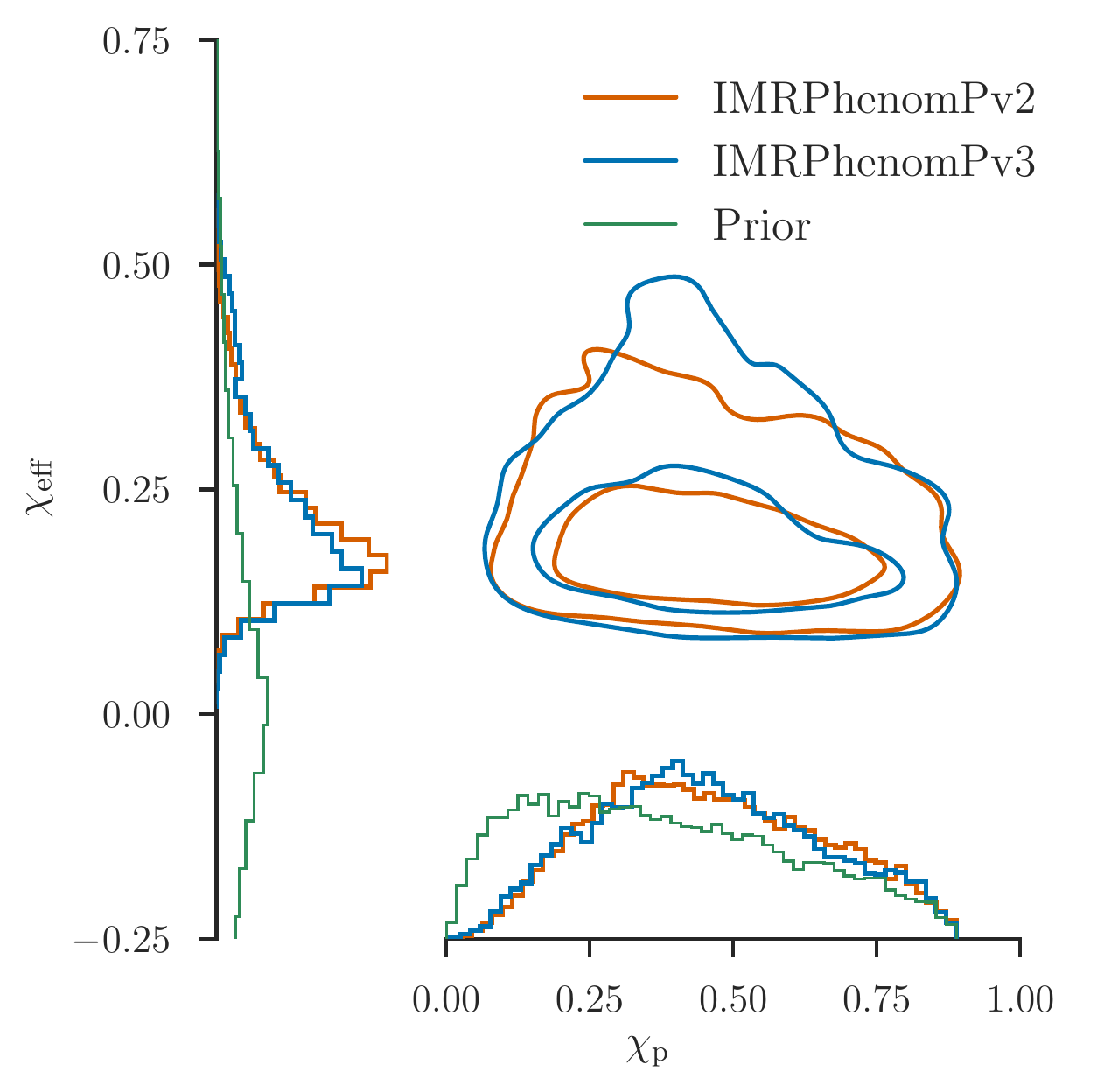}
    \includegraphics[width=\columnwidth]{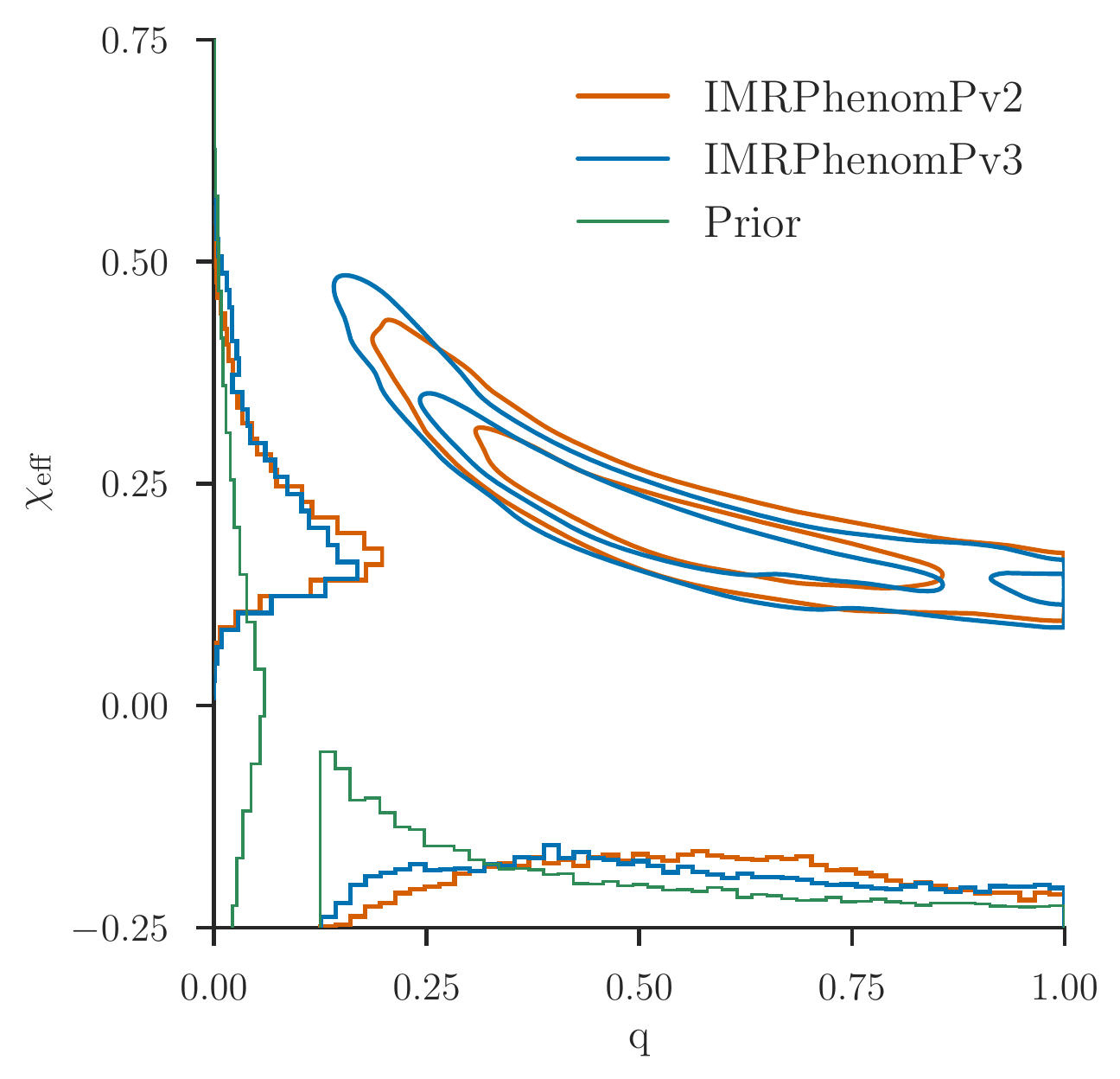}
 \caption{Posterior probability densities for $\chi_{\text{eff}}-\chi_p$ (left) and $\chi_{\text{eff}}-q$ (right) for GW151226 using {\tt IMRPhenomPv2} and {\tt IMRPhenomPv3}. Despite the more accurate description of 2-spin dynamics of {\tt IMRPhenomPv3} the posteriors are consistent, demonstrating the difficulty of measuring 2-spin dynamics. }
 \label{fig:151226_posteriors}
\end{figure*}

As a first application of the new {\tt IMRPhenomPv3} model we study whether the improved two-spin prescription allows for improved parameter extraction from existing GW signals. We focus on event GW151226 as it is the only system with strong support for at least one spinning BH~\cite{2016PhRvL.116x1103A,Vitale:2017cfs}. In spite of this, GW151226 has never been analyzed with a precessing model that employs two-spin dynamics: on the one hand, GW151226 has a low-enough total mass that an analysis with {\tt SEOBNRv3} is not computationally feasible; simultaneously, GW151226 is so massive that the merger phase of the coalescence is in band, making an analysis with precessing inspiral-only models, such as {\tt SpinTaylorT4}, incomplete.

The new {\tt IMRPhenomPv3} waveform model constructed here meets both criteria of computational efficiency and full coalescence description, allowing us to perform the first two-spin analysis of GW151226. We use the Bayesian Inference code {\tt LALInferenceNest}~\cite{2015PhRvD..91d2003V} and publicly available data from the LIGO Open Science Centre (LOSC, \href{https://losc.ligo.org}{\texttt{losc.ligo.org}}). We estimate the power spectral density noise using on-source data and the {\tt BayesWave} algorithm~\cite{Littenberg:2014oda,Cornish:2014kda}. We marginalise over detector calibration amplitude and phase uncertainty using values provided in~\cite{PhysRevX.6.041015}. We use a spin prior that is uniform in direction and magnitude up to $0.89$.
In what follows, we present and compare results obtained using {\tt IMRPhenomPv2} using a reduced-order-quadrature approximation to the likelihood~\cite{Smith:2016qas} and {\tt IMRPhenomPv3}.

Table~\ref{table:par-table} presents our results for the intrinsic parameters.
We quote the median value from the 1-D marginalised posterior and the associated
$90\%$ symmetric credible interval.
Noteworthy differences are a broader uncertainty in the primary mass
resulting in {\tt IMRPhenomPv3} favouring a slightly more asymmetric system,
although both models are within the statistical uncertainty of each other.
Overall we find consistent results between the two models as well as the published
\ac{LVC} analysis~\cite{PhysRevX.6.041015}.
We also have an estimate of the Bayes factor for a
coherent signal across the two LIGO detectors versus an incoherent signal or noise.
We find a slightly larger Bayes factor for {\tt IMRPhenomPv3} despite the extra
degrees of freedom from the full 2-spin description

Figure~\ref{fig:151226_posteriors} shows the 2-dimensional posterior densities for $\chi_{\text{eff}}-\chi_p$\footnote{Note that {\tt IMRPhenomPv3} no longer explicitly uses the effective precession parameter $\chi_p$. However we show it here for direct comparison to {\tt IMRPhenomPv2} results. } (left) and $\chi_{\text{eff}}-q$ (right) obtained from both waveform models as well as the 1-dimensional prior distributions. Both plots demonstrate broad agreement with {\tt IMRPhenomPv3} marginally favoring more unequal masses.
This result is also consistent based on the comparisons to \ac{NR} waveforms presented in the previous section.
The posterior for $\chi_p$ is consistent with its prior, confirming the absence of evidence of spin-precession in GW151226.
Overall we find that a re-analysis of the merger event GW151226 with our new waveform model does not provide new insights into the nature of the source but instead reinforces our current understanding of the source.

The consistency of the results implies that two-spin
precession effects do not impact this signal in agreement with previous studies predicting that aligned-two-spin effects are not easily measurable~\cite{Purrer:2015nkh}.
However, it is possible that certain binaries and orientations might make it possible to measure both spins simultaneously. This could be achieved, for example, with precessing models that also include higher multipoles modes, or with louder signals observed by more detectors.

\begin{table}[t]
\centering
  \begin{tabular}{lll}
    \hline \hline
Parameter                     & {\tt IMRPhenomPv2}       & {\tt IMRPhenomPv3}  \\ \hline
Primary Mass: $m_{1} (M_{\odot})$           & $13.68^{+7.93}_{-3.13}$  & $14.45^{+10.23}_{-3.96}$ \\
Secondar Mass: $m_{2} (M_{\odot})$           & $7.73^{+2.10}_{-2.41}$   & $7.35^{+2.57}_{-2.52}$ \\
Total Mass: $M_{\rm{tot}} (M_{\odot})$    & $23.37^{+6.00}_{-1.10}$  & $23.80^{+8.33}_{-1.55}$ \\
Mass ratio: $q$                           & $0.57^{+0.36}_{-0.32}$   & $0.51^{+0.44}_{-0.31}$ \\
Effective Spin: $\chi_{\rm{eff}}$             & $0.19^{+0.18}_{-0.07}$   & $0.20^{+0.23}_{-0.09}$ \\
Precession Parameter: $\chi_{\rm{p}}$               & $0.44^{+0.34}_{-0.28}$   & $0.44^{+0.35}_{-0.28}$ \\
Log Bayes Factor: $\rm{Log}(\mathcal{B})$       & 50.707                   & 51.291           \\ \hline \hline
\end{tabular}
\caption{Parameter Table for GW151226. Masses are defined in the source frame.
We quote the median and the 90\% symmetric credible interval of the 1D marginalised posterior distributions.}
\label{table:par-table}
\end{table}

\section{Discussion}

We have presented an upgrade to the
phenomenological model {\tt IMRPhenomPv2} called {\tt IMRPhenomPv3}.
This model predicts the \ac{GW} polarisations computed using the
the dominant $\ell = |m| = 2$ multipole in the co-precessing
frame, from non-eccentric merging \acp{BBH} with generically orientated spins.
Our upgrade consists of replacing the model for the precession dynamics of
\cite{Hannam:2013oca} that was derived under the assumption of a
single spin precessing \ac{BBH} system with the more accurate
analytic model from \cite{Chatziioannou:2017tdw,Chatziioannou:2016ezg}
that contains two-spin effects.

We have validated our new model against a large set of precessing
\ac{NR} waveforms. Although our selection of
\ac{NR} waveforms is biased towards mass-ratios $ < 3 $ and
spin magnitudes $ < 0.5 $ we find that all three models considered
generally perform well however, we have identified a clear region in parameter
space, mass-ratios $ > 3$ where all models begin to loose accuracy.
Encouragingly we find that {\tt IMRPhenomPv3} greatly outperforms the previous model
for the most extreme case we considered, with the largest improvement being $\sim 12\%$.
This improvement suggests that {\tt IMRPhenomPv3} can be utilised
in a much wider parameter space than {\tt IMRPhenomPv2} was found to be
reliable.
We emphasise the importance of \ac{NR}  to continue to push the
capabilities of precessing \ac{BBH} simulations that will allow more
stringent tests of waveforms models and ultimately lead to more accurate
waveform models being developed.

As a first application we re-analysed GW151226 with our two-spin
model and find consistent results compared with the single-spin model
thereby reinforcing our current inference on the nature of GW151226.

As our waveform model is analytic one can evaluate the model using
a non-uniform grid of frequencies. This is essential for methods such as
ROQ~\cite{Smith:2016qas} and the multi-banding
technique of~\cite{Vinciguerra:2017ngf}. These methods can increase the
computational efficiency of sampling-based parameter estimation by factors of 300 for
low mass systems.

Finally we note that there are a number of physical effects that are ignored.
We do not model any asymmetry between the positive and negative
$m$ modes which are responsible for large out-of-plane recoil kick velocities.
In the underlying, aligned-spin model we do not include
effects from higher order multipoles however, we have recently
developed a method to do this~\cite{PhysRevLett.120.161102}
and we are currently determining the
accuracy of a precessing model that includes higher order multipoles
in the co-precessing frame.

\begin{acknowledgments}
We thank Carl-Johan Haster for kindly supplying the PSD computed
using {\tt BayesWave} for the parameter estimation study of GW151226.
We thank Sylvain Marsat for useful discussions.
S.K. and F.O. acknowledge support by the Max Planck Society's Independent Research Group Grant.
We thank the Atlas cluster computing
team at AEI Hannover where this analysis was carried out.
M.H. was supported by Science and Technology Facilities Coun- cil (STFC) grant ST/L000962/1
and European Research Council Consolidator Grant 647839.
This work made use of numerous open source computational packages
such as python~\cite{python}, NumPy, SciPy~\cite{scipy}, Matplotlib~\cite{Hunter:2007}
and the \ac{GW} data analysis software library {\tt pycbc}~\cite{pycbc-software}.
\end{acknowledgments}

\bibliography{phenompv3}

\end{document}